\documentclass[article,10pt]{iopart}

\usepackage{iopams}
\usepackage[utf8]{inputenc}
\usepackage{amssymb}
\usepackage[english]{babel}
\usepackage{graphicx}
\usepackage{amsfonts}

\usepackage{cite}
\usepackage[printwatermark]{xwatermark}
\usepackage{xcolor}
\usepackage{graphicx}
\usepackage{lipsum}

\begin{document}

\title[Effect of atomic lensing]{Dispersive detection of atomic ensembles in the presence of strong lensing}

\author{A. B. Deb, J. Chung and N. Kj{\ae}rgaard}

\address{Department of Physics, QSO---Centre for Quantum Science, and Dodd-Walls Centre for Photonic and Quantum Technologies, University of Otago, Dunedin, New Zealand}
\ead{amita.deb@otago.ac.nz}
\vspace{10pt}

\begin{abstract}
We experimentally and theoretically investigate in-medium propagation effects of off-resonant light in dense, spatially homogeneous ultacold atomic gases. Focussing on frequency modulation spectroscopy as the dispersive detection tool of atoms, we observe that the refractive gradient-index lenses presented by localised atomic ensembles can significantly modify the interpretation of the dispersive signal even for large probe detuning, owing to the collective dispersive response of the atoms. We identify criteria for distinguishing between thin and thick atomic lenses, leading to either diffraction-dominated and lensing dominated regimes for the outgoing probe beams. Our findings are consistent with experimental data and solutions of paraxial wave equation for light propagation. Our study provides important practical insights for dispersive, minimally intrusive optical detection and imaging schemes of ultracold atoms and will be valuable for choosing optimal parameter regimes in numerous applications. 

\end{abstract}

%
%
%
%
%

\section{Introduction}

Strong atom-light interaction is the foundation of many emerging quantum technologies, such as quantum metrology beyond the standard quantum limit \cite{Hammerer2010,Appel2009,Kuzmich1997, Hosten2016,Napolitano2011}, continuous variable quantum communication via quantum non-demolition measurements \cite{Duan2001,Kuzmich2000} and quantum memories \cite{Fleischhauer2002}. A key parameter for strong atom-light coupling is the so-called cooperativity \cite{Tanji-Suzuki2011} which is a measure of the spatial mode overlap between the incident mode of light and the mode scattered by the atoms. High co-operativity is typically achieved by placing atoms in high-finesse resonant optical cavities. In recent years, strong atom-light interactions between single atoms and light in free-space and guided geometries have also been investigated using high-NA lenses \cite{Tey2009,Leuchs2013}, optical nanofibers \cite{Volz2014,Ostfeldt2017,Ruddell2017} and photonic crystal waveguides \cite{Burgers2019}.  There has also been a substantial interest in designer atomic arrays to cooperatively enhance light-atom coupling \cite{Bettles2016, Chomaz2012,Facchinetti2016,Shahmoon2017}. 

Ensembles of randomly positioned atoms dispersively interacting with a paraxial light beam can potentially give rise to high free-space cooperativity. This is owing to the fact that scattered light fields from individual atomic dipoles interfere predominantly constructively in the forward direction, strongly enhancing the overlap of the scattered mode with the incident mode. As it is impossible to tell which atom scattered a particular photon detected in the forward direction, the light field is quantum-entangled with the collective atomic coherences. As a result its detection leads to backaction on the atoms, leading to squeezing and quantum-enhanced measurement precision \cite{Windpassinger2009b}. 

Dispersive optical probing has been an important tool as a minimally destructive diagnostic for ultracold atomic samples. This constitutes the basis of `real-time' monitoring of the internal and external degrees of freedom of the trapped ultracold atoms \cite{Takahashi1999, Chaudhury2006, Petrov2006, Windpassinger2008, Windpassinger2008a, Kubasik2009, Kohnen2011,Bernon2011, Sawyer2012,Deb2013, Sawyer2017, Jammi2018}, as well as for minimally destructive imaging of atoms \cite{Andrews1996, Ketterle1999,Meppelink2009a} at or below the shot noise level \cite{Kristensen2017,Gajdacz2016}. This underpins many interesting feedback loop or optimal (open-loop) control procedures for stabilisation of atomic motion \cite{Wilson2007, Hush2013,Steck2003} and their quantum spin \cite{Vanderbruggen2013, Behbood2013, Anderson2015,Cox2016}

Ultracold trapped atomic samples typically realised in the lab have inhomogeneous density profiles and finite spatial extents. This renders the one-dimensional Beer-Lambert type description of ensemble-light coupling insufficient and motivates a three-dimensional description. The effect of the geometry of the atomic ensemble and paraxial light field on the efficiency of cooperative scattering was for example theoretically discussed by M{\"u}ller \textit{et.~al.} \cite{Mueller2005}. Along the same lines, Baragiola \textit{et.~al.} studied theoretically \cite{Baragiola2014} the effect of mode-matching between the atomic ensembles and the probe light field in the quantum regime in the context of spin-squeezing via continuous QND measurements. They interestingly concluded that spin-squeezing can even be enhanced in the presence of spatial inhomogeneity compared to homogeneous samples.
 
 Large resonant optical density ($OD_{res}$) is a key figure of merit for quantum-interfacing atomic ensembles with light. For spatially non-uniform samples of atoms, the distortion of an off-resonant light field during its propagation inside the sample can have a profound effect on the outgoing spatial mode of the light field. For example, in a cigar-shaped gaussian atomic sample with high on-axis optical depth, a significant transverse gradient of optical phase can develop when an off-resonant plane wave probe propagates through the sample. According to the Poynting theorem, the local propagation direction of light is bent along the direction of the phase gradient \cite{Matzliah2017}. Thus the probe light field is diverted out of the incident mode by the atomic sample, which in this case behaves as a gradient-index lens. Such lensing by micron scale cold atomic samples has been studied recently in \cite{Roof2014,Noaman2018}.
 
In this paper, we experimentally and theoretically study the effect of lensing in dispersive interfacing of atomic ensemble with paraxial light for a particular implementation of optical probing, based on frequency modulation spectroscopy (FMS) \cite{Bjorklund1980,Lye1999,Bernon2011}. FMS offers the advantages of a zero baseline and excellent passive common-mode phase noise reduction between the reference and signal arms of an effective Mach-Zehnder interferometer. We observe that the effect of lensing in inhomogeneous ultracold samples can be substantial even when the probe field is hundreds of linewidths away from the atomic resonance. While our study uses a specific implementation of dispersive probing, the results are relevant to a broad class of experiments probing atomic samples with off-resonant light. Such geometrical and in-medium propagation effects relevant in three-dimensional interfaces have not been studied before in the context of frequency modulation spectroscopy. Our studies are particularly important for atomic samples with large optical depths, small spatial size and strong density gradients, such as a Bose-Einstein condensate.
 
The paper is organised as follows. In section 2, we briefly introduce the basic light-atom interactions and how collective effects dictate the propagation of light fields in a medium consisting of dipoles. In section 3, we describe the macroscopic fields for homogeneous and inhomogeneous samples, derive an expression for the focal length atomic gradient index lens and the criterion where lensing plays a dominant role in dispersive probing.  In section 4, we introduce frequency modulation spectroscopy, and derive expressions for FMS signal in the case of a three-dimensional interface, as in our experiment. In section 5, we discuss our experimental setup, present our experimental observations and interpret the data based on our analytical model as well as a paraxial wave equation model. In section 7, we summarise our results and provide an outlook on the implications of our findings and future studies.

\section{Basic problem} 


When a ground state atom is placed in a propagating light field, which has its frequency in the neighbourhood of an atomic transition to an excited state, light will be scattered. For a sufficiently low intensity the scattered light will have the same frequency as the incoming light, it will have the same spectrum as the incoming light (i.e., it will not reflect the natural linewidth of the excited state), and the two fields will be mutually coherent \cite{CohenTannoudji2004,Carmichael2013}. The scattered light will, however, generally be in different spatial mode to the incoming field. In a typical experimental configuration, the incoming light would be in the form of a Gaussian laser beam and light would be scattered into a dipole radiation pattern. For example, a circularly polarised incoming field propagating along the $z$ axis driving a circularly oscillating (rotating) electric dipole, the emitted field will be spherical wave with its power distributed as $\propto (1+\cos^2\theta)$ \cite{Corney1977}. Since the incoming and scattered fields are coherent they will interfere \cite{Aljunid2009}

In a semiclassical treatment, the steady-state dipole moment of a two-level atom perturbed by a weak, near resonant light field, which is $E(t)=E_0\cos\omega t$ at the position of the atom, is found to be \cite{Knight1983}
\begin{equation}\label{eq:dipole}
  D(t)=\frac{\mu^2/\hbar}{\sqrt{\Delta^2+\Gamma^2/4}}E_0\cos(\omega t-\eta),
\end{equation}
where $\eta=\arctan(\Gamma/2\Delta)$ is the phase lag of the dipole with respect to the driving electric field and $\mu$ is the magnitude of the electric (transition) dipole matrix element corresponding to the transition which is the related to the natural linewidth $\Gamma$ of the transition by $\Gamma = k^3 \mu^2/(3 \pi \epsilon_0\hbar)$. Here $k = 2\pi/\lambda$, where $\lambda$ is the wavelength of the transition. In particular, we note that on resonance $\Delta=0$ the dipole lags the driving field by $\pi/2$. For the oscillating induced dipole, the radiated electric field has a magnitude at the point $\mathbf{r} = (r,\theta,\phi)$ given by 
\begin{equation}\label{eq:dipoleEfield}
  E(\mathbf{r}) = \frac{D(t)}{8\pi\epsilon_0}(1 + \cos^2\theta)\frac{k^2}{r}
\end{equation}

In the case of an ensemble of atomic dipoles inside the volume $V'$ illuminated by an incident field $\mathbf{E}_{\mathrm{inc}}(\mathbf{r},t)$ and scattering coherently, the situation is complicated by the fact that each dipole sees the not only the incident field but also the dipolar scattered field owing to all other dipoles present. The total field is then given by \cite{Lalor1969, Foldy1945}  

\begin{equation}\label{eq:dipole_ensemble}
 \mathbf{E}(\mathbf{r},t) = \mathbf{E}_{\mathrm{inc}}(\mathbf{r},t) + \int_{V'} \nabla \times \left(\nabla \times \frac{\rho(\mathbf{r'})\alpha\mathbf{E}(\mathbf{r'},t-R/c)}{R}\right)dV'
\end{equation}
where $\rho$ is the local number density of the scatterers, $\alpha$ is the polarizability and $R = | \mathbf{r}- \mathbf{r'}|$ is the distance between a scatterer and the observation point. The polarizability of a two-level atom can be easily derived from Equation \ref{eq:dipole}, by writing $D(t)$ as the real part of a complex number, to be 
\begin{equation}\label{eq:alpha}
  \alpha = -\frac{\mu^2}{\hbar}\frac{1}{\Delta + i\Gamma/2}.
\end{equation}
Note that the total field $\mathbf{E}$ appears on both sides of Equation \ref{eq:dipole_ensemble}, so the integral equation must be solved self-consistently to obtain the total field. 

For a dilute ensemble of dipoles, where the total scattered field from all dipoles at the location of a dipole is small compared to the incident field, one can make the simplifying assumption that all dipoles are illuminated by the incident light field and scatter independently, giving rise to a resultant electric field that then interferes with the incident light field at the observation point. This approximation is frequently made for dispersive probing of ultracold atoms \cite{Mueller2005,Baragiola2014}. This reduces the problem of light scattering from an atomic ensemble to that of diffraction of the light field by the medium. Notably, if the geometry of the atomic ensemble is fixed then the scattered light field in the forward direction is cooperatively enhanced by a factor of $N$, where $N$ is the number of dipoles in the ensemble. This linear dependence of the dispersive signal on the number of scatterers forms the basis of quantum non-demolition measurement of collective atomic variables using dispersive probing. 

There are, however, common practical situations where the dipoles are not only driven by $\mathbf{E}_{\mathrm{inc}}$. A trivial example is a system of dipoles with a large physical extent $l$ along incident light field axis such that $l> 1/n\sigma$, where $n$ is the atomic density and $\sigma$ is the light scattering cross-section. In this case the incident field will be significantly extinguished within the medium due to the large optical depth of the medium. In typical dispersive probing experiments with ultracold atoms one uses atomic samples with an off-resonant optical depth $\ll$ 1, but even in this case, if the sample is inhomogeneous in the transverse direction, it can significantly alter the transverse phase profile of the electric field \textit{within} the medium. Such gradient-index lensing effect can significantly affect the interpretation of the dispersive signal, as we show here.

As we will show below, two regimes for lensing exist. In the first regime, the atomic cloud acts as a thin lens where the focal length is much larger than the axial dimension of the sample, and the linearity between the dispersive signal and the atom numbers hold true. In the other regime where the focal length is comparable or smaller than the axial dimension, the linearity breaks down. Hence the relevant parameter is the lens Fresnel number $\mathcal{F} = a^2/f \lambda $, where $a$ is the transverse dimension of the sample and $f$ is the focal length. In regime corresponding to $\mathcal{F}  \ll 1$, a thin atomic lens is realized.


\section{Propagation of light in atomic gases}
 
\subsection{Homogeneous ensembles}

\begin{figure}
\centering
 \includegraphics[width=0.6\textwidth]{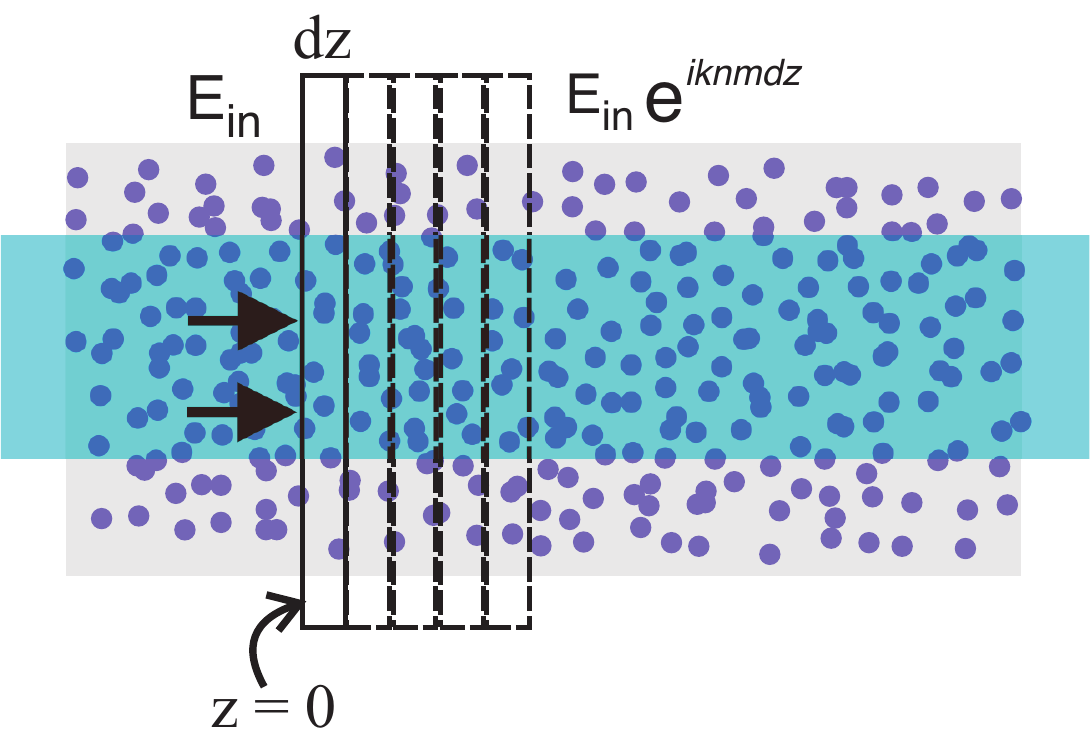}
\caption{The total field for scatterers inside a thin slice of thickness $dz$ at $z = 0$ is determined by the incident field and all other scatterers in the medium both $z>0$ and $z<0$. The Ewald-Oseen theorem states that the total field is a sum of two components, one that cancels the 
incident field and the other propagates through the medium with a wavevector $nk$, where $n$ is refractive index. }
\label{LensingFig1}
\end{figure}

To set the scene, we first consider a homogeneous dilute ensemble of atomic scatterers. Since these are discrete point scatterers, there will be randomness present at some level, so we assume that only the average density of atoms is constant and the positions of the scatterers are otherwise random. The fluctuations of atomic positions lead to incoherent (diffuse) scattering in non-forward directions \cite{Baragiola2014}. In a truly continuous medium, as is well-known \cite{Fearn1996a}, such diffuse scattering is absent and the description of wave propagation is strictly defined by refraction, reflection and transmission. To model the propagation of light in a homogeneous medium with discrete random scatterers with a mean inter-particle distance $D$, we consider a plane wave light field polarized in the x-direction, given by $\mathbf{E}_{\mathrm{inc}} = \mathbf{\hat{x}}E_{\mathrm{inc}} e^{i(kz-\omega t)}$ be incident in the medium and focus on a thin slice at $z = 0$ of width $dz \gg D \gg 1/k$ (Figure \ref{LensingFig1}). The total field driving the dipoles inside this slab is the sum of the external incident field and the dipolar field coming from all dipoles inside the medium - both the forward-propagating scattered field owing to the slab $z<0$ and backward-propagating field from slabs at $z>0$. According to the Ewald-Oseen extinction theorem of classical optics \cite{Fearn1996a, Newton1982}, the total scattered field at any point inside the medium can be decomposed into two parts: one that exactly cancels the external field and the other that propagates in the forward direction with an amplitude $2E_{\mathrm{inc}}/(n+1)$ and a phase $e^{i(nkz -\omega t)}$, where $n$ is the refractive index given by 
\begin{equation}\label{eq:refractiveindex}
 n = \sqrt{1+\frac{1}{\epsilon_0}\rho\alpha}.
\end{equation}
This is a remarkable result in that even though most of the space between the scatterers is empty, the medium acts has having a continuous refractive index determined by the \textit{average} density and this holds no matter how dilute the system is \cite{Fearn1996a}. Making the assumption that
\begin{equation}\label{eq:refractiveindex1}
 |n - 1| \ll 1
\end{equation}\label{eq.RI1}
we can write the field right after the first slab as  $\mathbf{E}(\delta z) = \mathbf{E}(0) e^{ikn\delta z}$ which is the effective field driving the dipoles in the following slab. The field after passing through $m$ such slabs so that $\mathit{l} = m\delta z$ the field is $\mathbf{E}(z = \mathit{l}) = \mathbf{E}(0) e^{iknm\delta z} = \mathbf{E}(0) e^{iknl}$. The assumption in Equation \ref{eq:refractiveindex1} also allows us to expand Equation \ref{eq:refractiveindex} as $n \simeq 1 + \frac{\rho\alpha}{2\epsilon_0}$, such that,
\begin{equation}\label{eq:dipole2}
\mathbf{E}(z =\mathit{l}) =  \mathbf{E}(0) e^{ik(1 + \frac{\rho\alpha}{2\epsilon_o})\mathit{l}} =  \mathbf{E}(0) e^{ik\mathit{l}} e^{\frac{ik\rho\alpha\mathit{l}}{2\epsilon_o}}.
\end{equation}
With the further assumption that 
\begin{equation}\label{eq:refractiveindex1b}
\frac{k\rho|\alpha| l}{2\epsilon_o} \ll 1,
\end{equation}
Equation \ref{eq:dipole2} reduces to 
\begin{equation}\label{eq:refractiveindex2k}
\mathbf{E}(z) \simeq  \mathbf{E}(0)e^{ikz}\left(1+  i\frac{k\rho\alpha z}{2\epsilon_o}\right).
\end{equation}
In this regime, the scattered light field in the forward direction (the second term in Equation \ref{eq:refractiveindex2k}) is a linear function of the total number of atoms (for a fixed sample volume), as is frequently required in QND experiments \cite{Tanji-Suzuki2011,Vanderbruggen2013,Cox2016,Napolitano2011,Hosten2016}. For instance, assuming an ensemble with a density $\rho = 2 \times 10^{19}$\,m$^{-3}$, $\Delta = 50 \Gamma$ and $\lambda = 780$\,nm, one must have $\mathit{l}\ll 30\,\mu$m to fulfil condition~\ref{eq:refractiveindex1}. In the following we consider the regimes where this assumption breaks down, as well as the assumption of homogeneity.

\subsection{Transversely inhomogeneous ensembles}\label{GRIN_theory}

\begin{figure}
\centering
 \includegraphics[width=0.74\textwidth]{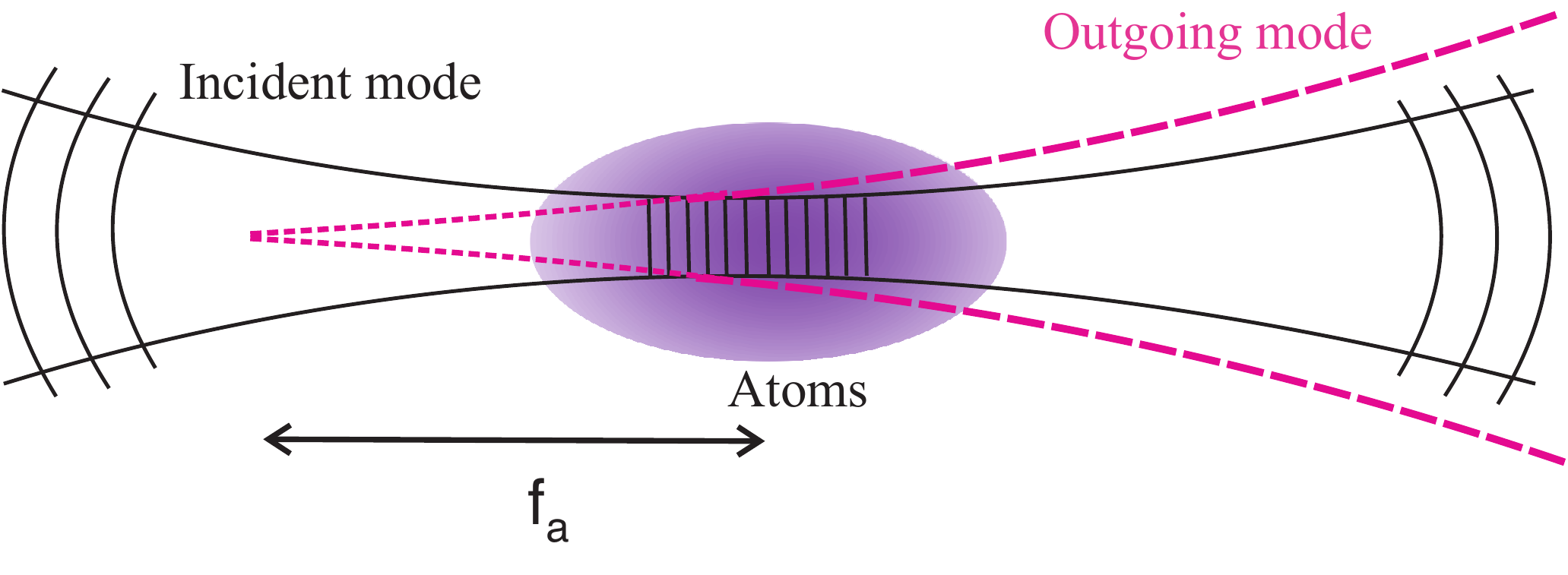}
\caption{Transversely inhomogeneous atomic samples act as a gradient index lens to the incoming light. The focal length of this lens is negative for blue-detuned light. }
 \label{LensingFig2}
\end{figure}

Trapped atomic ensembles, in practice, often presents a spatially localised sample typically with substantial density gradients both in the radial and axial directions with respect to a probe light beam. Let us focus on the radial dimension first. An atomic cloud with a finite radial extent acts as a diffractive aperture to the incoming beam. Moreover, any refractive index gradient causes the incoming beam to acquire a spatially varying phase shift, causing an effective lensing similar to a gradient-index (GRIN) lens (Figure \ref{LensingFig2}). Assuming a gaussian radial density distribution of atoms $\rho(r) = \rho_0 \exp({-r^2/\sigma_r^2})$ and substituting into Eq. \ref{eq:dipole2}, we obtain the field after passing through a slab of length $\delta z$ through the sample:
\begin{equation*}\label{eq:lens1}
\mathbf{E}(z =\delta z) =  \mathbf{E}(0) e^{ik(1 + \frac{\rho\alpha}{2\epsilon_o})\delta z} = \mathbf{E}(0) e^{ik\delta z} e^{\frac{ik\alpha\delta z}{2\epsilon_o}\rho_o\exp{(-\frac{r^2}{\sigma_r^2})}}.
\end{equation*}
Taking only into account the radial extent $r \lesssim \sigma_r $, such that $\rho(r) \simeq \rho_o(1 - r^2/\sigma_r^2)$,
\begin{equation}\label{eq:lens2}
\mathbf{E}(z =\delta z) =  \left \{\mathbf{E}(0) e^{ik\delta z} e^{\frac{ik\alpha \rho_o}{2\epsilon_o}\delta z}\right\} e^{-\frac{ik\alpha \rho_o\delta z\, r^2}{2\epsilon_o\sigma_r^2}}.
\end{equation}
This equation clearly shows that the effect of the atomic cloud is to convert a plane wavefront to a paraxial-spherical one, just as a thin lens would do. Comparing Eq. \ref{eq:lens2} to the phase transformation of a thin lens with a focal length $f$ \cite{Adams2019book}
\begin{equation}\label{eq:lens3}
E  \rightarrow E e^{-\frac{ikr^2}{2f}},
\end{equation}
we obtain an expression for the focal length of the atomic gradient-index lens:
\begin{equation}\label{eq:lens4}
f_a = \frac{\epsilon_o\sigma_r^2}{\alpha\rho_o \delta z}.
\end{equation}
Equation \ref{eq:lens4} reduces to the widely used formula the focal length of a ball lens \cite{Andrews1996,Roof2014} in the case $\sigma_r = \delta z$. We note that for a trapped Bose-Einstein condensate with a Thomas-Fermi (inverted parabolic) radial distribution, the expression is exact. As the light propagates through more and more slabs, the effect of diffraction and lensing has to be taken into account concurrently for each individual slab. For long samples and high atomic densities, the sample will cease to be a thin lens, as light bends significantly \textit{within} the medium. The expression for the focal length given by Equation \ref{eq:lens4} stays valid for a thick lens as long as the paraxial approximation is valid, and the phase factor in the term inside the curly bracket of Equation \ref{eq:lens2} is less than $\pi$ \cite{Adams2019book}:
 \begin{equation}\label{eq:lens6}
\frac{k\alpha \rho_o}{2\epsilon_o}\delta z \lesssim \pi,
\end{equation}
which is equivalent to requiring $\mathcal{F} \lesssim 1$, where the \textit{lens Fresnel number} $\mathcal{F}$ is defined by
 \begin{equation}\label{eq:lens5}
\mathcal{F} = \frac{\sigma_r^2}{\lambda f_a}.
\end{equation}
The Lens Fresnel number thus provides a qualitative measure for distinguishing between the thin and thick lens regimes. The left-hand side of Equation \ref{eq:lens6} is the ``peak phase shift" in the one-dimensional picture, as evident from Equation \ref{eq:refractiveindex2k}, and thus a peak phase-shift of $\pi$ deleneates between the thin and thick lens regimes.

\section{Heterodyne detection of dispersive response: phase modulation spectroscopy} 

Phase modulation spectroscopy (PMS) was developed in the 1980's as an ultra-sensitive probe for small absorption and phase shifts of light \cite{Bjorklund1980a,Bjorklund1983}. For this, a single-frequency laser source is phase modulated electro-optically at microwave frequencies and the resultant light beam has frequency sidebands imprinted on it. For weak modulation, only the first order sidebands are prominent, and phase modulation is equivalent to frequency modulation \cite{Supplee1994} (the reason why the technique is also sometimes knows are frequency modulation spectroscopy). The microwave frequency is chosen such that one of the sidebands (the \textit{probing sideband}) is close to an optical resonance of the atoms and the carrier and all other sidebands are far-detuned from the resonance.  In the absence of the atoms, the beat notes arising from the interference of the carrier and the positive and negative frequency sidebands cancel each other, resulting in a null signal at the heterodyne frequency. When the atoms are present, the probing sideband incurs both extinction and phase shifts, which results in a non-zero signal at the modulation frequency. More recently, the technique has been employed to perform minimally destructive dispersive probing of trapped ultracold samples to probe fast dynamical processes in-situ \cite{Lye1999,Bernon2011,Deb2013,Sawyer2017,Vanderbruggen2013,Sawyer2018,Sawyer2012}. Phase modulation spectroscopy offers several important benefits, passively phase-stable interference arms, excellent common-mode phase noise and low-frequency noise (such as flicker noise) reduction and its relatively simple implementation.  


\subsection{1D interface: plane wave, homogeneous samples}\label{sec:41}
Let us assume a plane wave $\mathbf{E}_{0} = \mathbf{\tilde{E}}_{0}e^{-i\omega t}$ that is electro-optically phase-modulated by a microwave field with frequency $\Omega$, here $\mathbf{\tilde{E}}_{0} = E_{0}\hat{\mathbf{x}} e^{ikz}$. For a sufficiently small modulation depth $m$, the resultant field to a good approximation is given by 
 \begin{equation}\label{eq:PMS1}
\mathbf{E}_{in} = \mathbf{\tilde{E}}_{c}e^{-i\omega t} + \mathbf{\tilde{E}}_{b}e^{-i(\omega + \Omega) t} - \mathbf{\tilde{E}}_{r}e^{-i(\omega - \Omega) t}.
\end{equation}
Here the subscripts `c', `b' and `r' stand for carrier frequency and blue (higher frequency) and red (lower frequency) sidebands and $|E_b| = |E_r| = \frac{m}{2}|E_c|$. The minus sign in front of the red sideband term reflects an additional phase of $\pi$ that arises from phase-modulation. When this beam passes through a homogeneous ensemble, the probing sideband (which we assume to the red one) acquires a (complex) phase shift $\delta = \delta_i + i\delta_r$, such that $\mathbf{\tilde{E}}_{r} = E_{r}\hat{\mathbf{x}} e^{ikz}e^{i\delta}$ (see Eq. \ref{eq:dipole2}), while the carrier and the blue sidebands leave the sample unchanged. The outgoing wave
\begin{equation}\label{eq:PMS2}
 \mathbf{E}_{out} = \mathbf{\tilde{E}}_{c}e^{-i\omega t} + \mathbf{\tilde{E}}_{b}e^{-i(\omega + \Omega) t} - \mathbf{\tilde{E}}_{r}e^{-i(\omega - \Omega) t} e^{i\delta}
\end{equation}
is detected on a fast photodiode which measures the resultant intensity
\begin{equation}\label{eq:PMS3}
I = \frac{1}{2}c\epsilon_o  \mathbf{E}^{*}_{out}.\mathbf{E}_{out}.
\end{equation}
Filtering out the dc and the $2\Omega$ components from the photodetector signal, one obtains the FMS signal 
\begin{equation}\label{eq:PMS4}
\mathcal{S}(\Omega) = \frac{1}{4} m \mathcal{A} \kappa c\epsilon_o  \left\{ |E_c|^2(1 - e^{-i\delta})e^{-i\Omega t} + c.c. \right\}.
\end{equation}
Here $\kappa$ is the photodetector responsivity and $\mathcal{A}$ is the detector area. For small phase shifts, $|\delta| \ll 1$, the amplitude of the sinusoidal PMS signal is proportional to the phase shift:
\begin{equation}\label{eq:PMS4}
\mathcal{S}(\Omega) = \frac{1}{2} m  \mathcal{A} \kappa c\epsilon_o  |E_c|^2 \cos(\Omega t)(i\delta_i - \delta_r).
\end{equation}
The quadrature and the in-phase parts of the signal, therefore, measure the absorption and the (real) phase shift respectively - yielding information about the optical response of the sample.

\begin{figure}
\centering
 \includegraphics[width=0.95\textwidth]{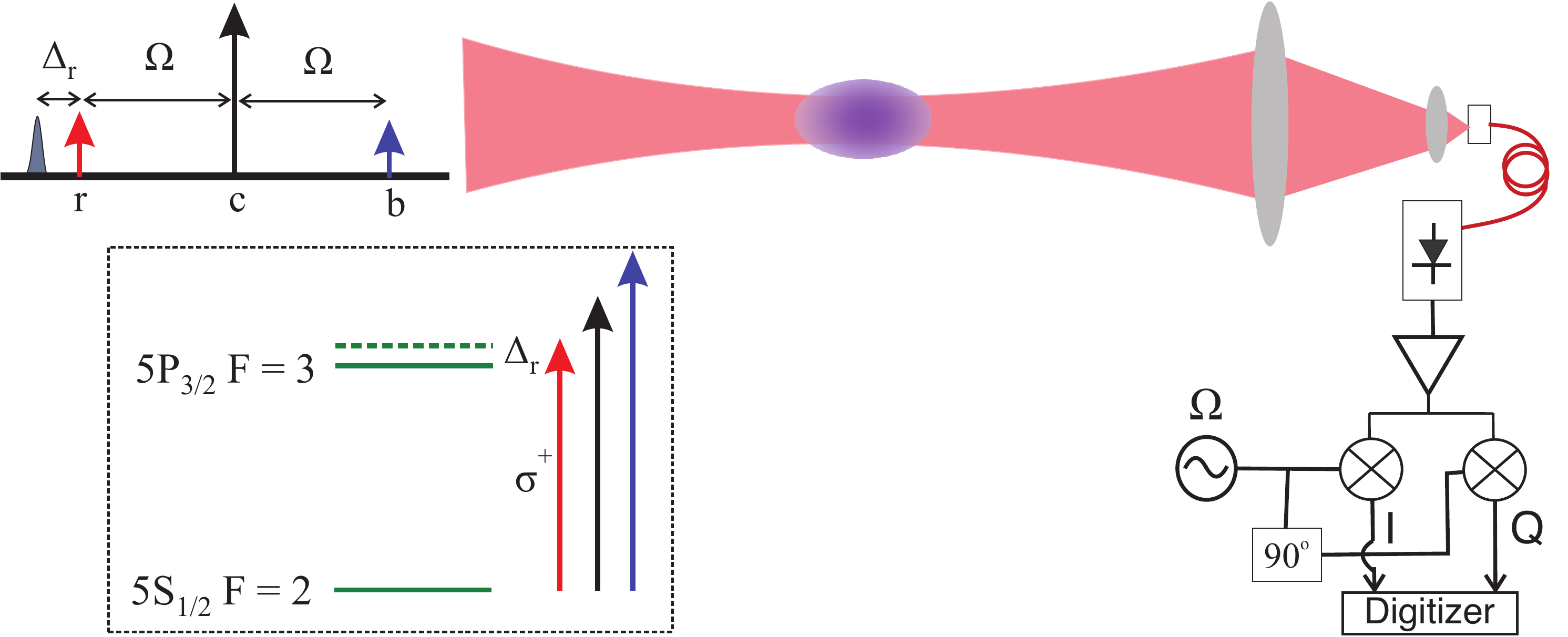}
\caption{The experimental setup. }
\label{LensingFig3}
\end{figure}

\subsection{3D interface: focussed Gaussian beam and trapped inhomogeneous ensembles}\label{sec:42}
For atomic ensembles with small spatial extent, it is necessary to focus the probe light beam to achieve high overlap of the light beam and the atomic distribution (Figure~\ref{LensingFig3}). The incident light beam is typically derived from a TEM$_{00}$ mode of a laser, the spatial terms of the electric field of which is given by
\begin{equation}\label{GB1}
  \mathbf{\widetilde{E}}(\mathbf{r}) = \left[\hat{\mathbf{e}}\mathrm{E}_{\mathrm{0}} \frac{w_0}{w(z)}\exp\{-\frac{r^2}{w^2(z)}\}\large \right ]\exp\{-i[kz - G(z)+\frac{kr^2}{2R(z)}]\}
\end{equation}
where  $\hat{\mathbf{e}}$ is the polarization of the light, $w_0$ is the beam waist, $w(z)~=~w_0[1+(z/z_R)^2]^{1/2}$, the radius of curvature $R(z) = z[1+(z_R/z)^2]$ and Gouy phase $G(z) = \arctan(z/z_R)$, where $z_R = \pi w_0^2/\lambda$ is the Rayleigh range. As in section \ref{sec:41}, the incident beam is then given by $
\mathbf{E}_{in}(\mathbf{r}) = \mathbf{\tilde{E}}_{c}(\mathbf{r}) e^{-i\omega t} + \mathbf{\tilde{E}}_{b}(\mathbf{r}) e^{-i(\omega + \Omega) t} - \mathbf{\tilde{E}}_{r}(\mathbf{r}) e^{-i(\omega - \Omega) t}$ ,
where $\mathbf{E}_{c,b,r}(\mathbf{r})$ all have the same spatial (Gaussian) mode as defined in Equation \ref{GB1}. Assuming again that the carrier and the blue sideband interact negligibly with the atoms, the outgoing beam is given by
\begin{equation}\label{eq:PMS21}
\mathbf{E}_{out}(\mathbf{r}) = \mathbf{\tilde{E}}_{c}(\mathbf{r}) e^{-i\omega t} + \mathbf{\tilde{E}}_{b}(\mathbf{r}) e^{-i(\omega + \Omega) t} - \{\mathbf{\tilde{E}}_{r} (\mathbf{r})+\mathbf{\tilde{E'}}_{r} (\mathbf{r})\} e^{-i(\omega - \Omega) t}.
\end{equation}
Here 
\begin{equation}\label{eq:PMS22}
\mathbf{\tilde{E'}}_{r} (\mathbf{r}) = \mathbf{\tilde{E}}_{r,tot} (\mathbf{r}) - \mathbf{\tilde{E}}_{r} (\mathbf{r}),
\end{equation}
 is the spatial part of the total scattered field, and $\mathbf{\tilde{E}}_{r,tot} (\mathbf{r})$ is the spatial part of the total field for red sideband in the presence of the atoms. The intensity at a detection point $\mathbf{r}$ in the far field is then given by
 \begin{equation}\label{11}
 I_\Omega(\mathbf{r},t) = \frac{1}{2}c\epsilon_0 \left [ \mathbf{\tilde{E}}_{\mathrm{c}}^{*}(\mathbf{r})\mathbf{\tilde{E}'}_{\mathrm{r}}(\mathbf{r})\exp\{i\Omega t\} + \mathrm{c.c.} \right]
 \end{equation}
and the PMS heterodyne signal is given by
\begin{equation}\label{12}
 S(\Omega) = \frac{1}{2}c\kappa\epsilon_0\left[\exp\{i\Omega t\} \int_\mathcal{A} d^3\mathbf{r} \mathbf{\tilde{E}}_{\mathrm{c}}^{*}(\mathbf{r})\mathbf{\tilde{E}'}_{\mathrm{r}}(\mathbf{r}) + \mathrm{c.c.} \right],
\end{equation}
where $\mathcal{A}$ is the effective aperture of the system, and $\kappa$ is the responsivity of the photodetector. The signal measures the overlap of the scattered mode with the incident reference mode (carrier), or equivalently by Equation \ref{eq:PMS22}, any deviation of the probe (red) sideband from the incident mode. It is thus sensitive to extinction, phase shift or any propagation effects such as lensing, of the red sideband due the presence of the atoms. 

\subsection{FMS signal detection at the baseband}\label{sec:43} 
The PMS spectroscopy signal is typically in the microwave domain ($\Omega > 1$\,GHz). It is therefore convenient to down-convert it to the baseband by mixing the signal with a reference microwave signal (local oscillator (LO)) that is phase-coherent to the signal employed for phase-modulation: $S_{ref,I} = V_0\cos(\Omega t + \theta) = (V_0/2)(e^{i(\Omega t + \theta)} + \mathrm{c.c.})$. The baseband signal is then given by
\begin{equation}\label{PMS14}
\mathcal{B}_I = \frac{1}{4}c\kappa\epsilon_0 V_0 \cos\theta \, \mathrm{Re} \{\int_\mathcal{A} d^3\mathbf{r} \mathbf{\tilde{E}}_{\mathrm{c}}^{*}(\mathbf{r})\mathbf{\tilde{E}'}_{\mathrm{r}}(\mathbf{r})\}
\end{equation}
The signal in Equation \ref{PMS14} can be maximised by adjusting the LO phase such that $\theta = 0$. In our implementation, we use IQ demodulation where we generate a second signal by mixing the heterodyne signal with $S_{ref,Q} = V_0\sin(\Omega t + \theta) = (V_0/2i)(e^{i(\Omega t + \theta)} - \mathrm{c.c.})$, resulting in the baseband signal
\begin{equation}\label{PMS15}
\mathcal{B}_Q = \frac{1}{4}c\kappa\epsilon_0 V_0 \sin\theta \, \mathrm{Im} \{\int_\mathcal{A} d^3\mathbf{r} \mathbf{\tilde{E}}_{\mathrm{c}}^{*}(\mathbf{r})\mathbf{\tilde{E}'}_{\mathrm{r}}(\mathbf{r})\}.
\end{equation}
From Equation \ref{PMS14} and \ref{PMS15}, we eliminate $\theta$:
\begin{equation}\label{PMS15}
\mathcal{B}= \sqrt{\mathcal{B}_I^2 + \mathcal{B}_Q^2 } = \frac{1}{4}c\kappa\epsilon_0 V_0 \left |\int_\mathcal{A} d^3\mathbf{r} \mathbf{\tilde{E}}_{\mathrm{c}}^{*}(\mathbf{r})\mathbf{\tilde{E}'}_{\mathrm{r}}(\mathbf{r})\right |.
\end{equation}

\section{Experiments}
\subsection{Experimental setup}\label{sec:51} 
We perform experiments with ultracold samples of $^{87}$Rb atoms in the $|F = 2,~m_F= 2\rangle$ spin-state trapped in an Ioffe-Pritchard (IP) magnetic trap with trapping frequencies $\omega_x = \omega_y = 2\pi\times$\,166\,Hz and $\omega_z = 2\pi\times 17$\,Hz, loaded from a magneto-optical trap (MOT) using a movable quadruple trap. The details of our setup have been reported elsewhere \cite{Deb2013,Sawyer2017}. For ultracold temperatures considered here, the atoms reside close to the trap minimum where the magnetic field direction is predominantly in the z-direction, which we choose as the quantisation axis. We control the number of atoms in the samples by adjusting the MOT loading while their temperatures are controlled by adjusting the final point of the radio-frequency ramp used for evaporative cooling. The setup for phase-modulation spectroscopy is shown schematically in Figure 3. In short, a master laser is frequency-locked to the $F = 2 \rightarrow F' = 3$ atomic transition of $^{87}$Rb using saturation absorption spectroscopy.  The laser light used as the carrier for dispersive probing is derived from a narrow-linewidth commercial diode laser and its frequency is stabilised using an offset beat-note lock using light from the master laser as reference. The probe laser can be locked to up to 7 GHz away from the resonance on either side.

We set the carrier frequency to be on the blue (higher frequency) side of the $F = 2 \rightarrow F' = 3$ resonance. The light passes through an fiber-coupled electro-optic modulator which is driven with a microwave frequency $\Omega = 2\pi\times3.7$\,GHz, and at a low modulation depth such that the first-order frequency sidebands contain about 5\% of the total power each. In this way, the red sideband is placed close to the $F = 2 \rightarrow F' = 3$ resonance with a detuning $\Delta_r$ that is adjusted by changing the offset-lock frequency. The light is focussed to a $1/e^2$ waist of $36\,\mu$m onto the atomic sample and is collected with a fast photodetector following an optical system with effective numerical aperture of 0.025 and with a collection efficiency of 75\% as measured without any atoms present. The photodetector signal is amplified before being demodulated using an IQ-mixer and sampled on a digitizer. The probe beam triplet is $\sigma^{+}$ polarized and has total power of 20\,$\mu$W before the atoms which we illuminate for a duration of 5\,$\mu$s.  For $\Delta_r = 2\pi\times 300$\,MHz, the smallest detuning used in this work, this corresponds to about 0.25 photon spontaneously scattered per atom per pulse. Due to the cycling nature of the $\sigma$ transition, optical pumping to other ground spin-states is negligible. 
The samples are subsequently imaged using resonant absorption imaging to deduce atom number. 

The in-phase and quadrature components of the signal recorded with the digitizer is processed using Equation \ref{PMS15} and integrated over the length of the pulse to obtain the dispersive signal.

\subsection{Experimental observations and interpretation}\label{sec:52}

\begin{center}
\begin{figure}
        \includegraphics[width=1.0\textwidth]{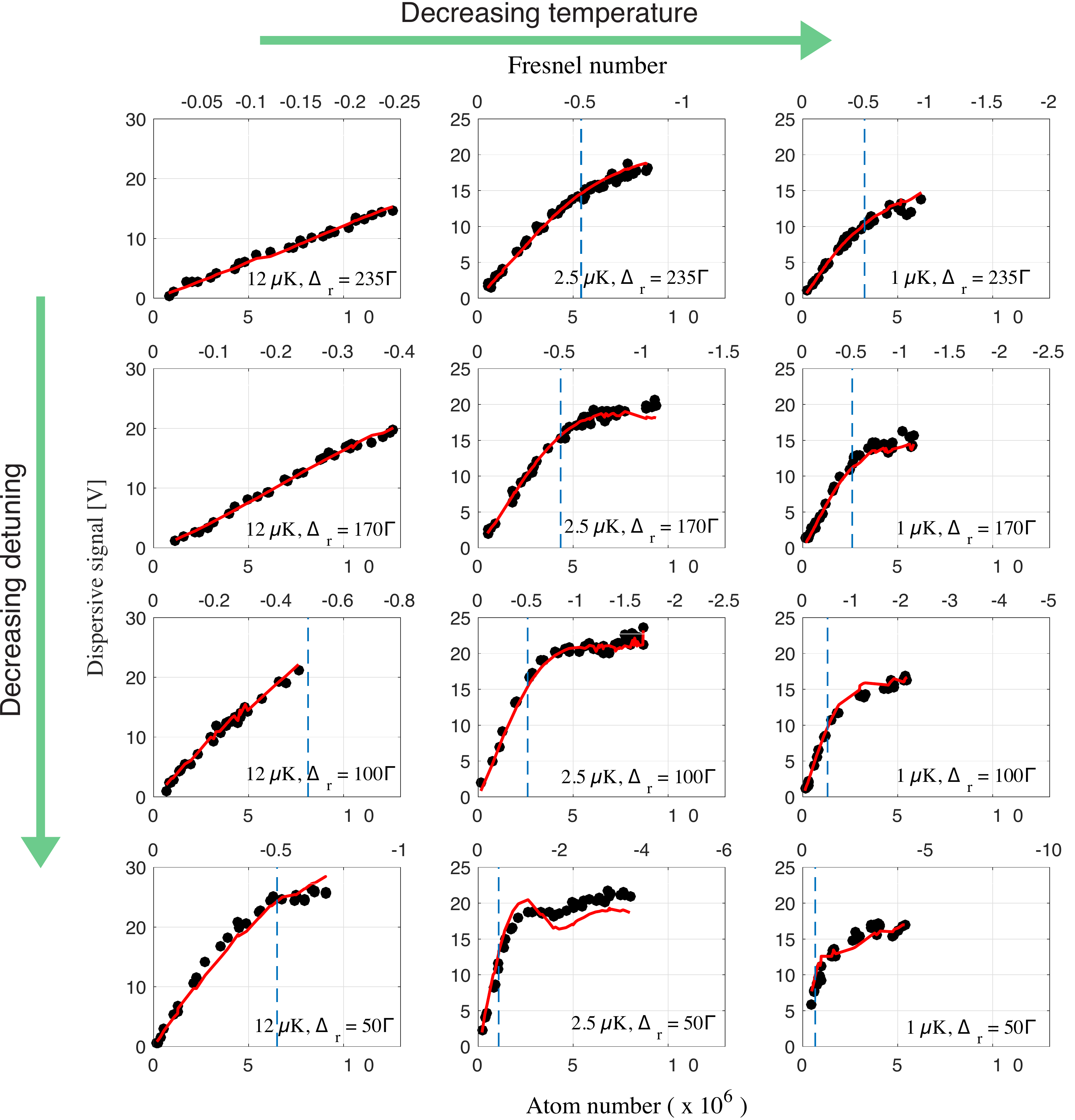}
     
        \caption{Dispersive signal as a function of atom number for temperatures for a range of probe detunings. Sample temperature decreases from left to right and the red sideband detuning of the probe changes from top to bottom. The bottom horizontal axis on each plot shows the atom number whereas the top horizontal axis shows the Fresnel number $\mathcal{F}$ based on Equation \ref{eq:lens4} and \ref{eq:lens4}. The black circles indicate the measured signals. The solid lines show the results of the simulations based on a paraxial wave equation. The vertical dashed lines indicate $\mathcal{F} = 0.5$ - distinguishing the lensing-dominated regime to its right from the diffraction-dominated regime to its left.}
        \label{LensingFig4}
\end{figure}
\end{center}

Figure \ref{LensingFig4} shows the dispersive signal as a function of atom number for atomic samples at temperatures T~=~1 ($\pm 0.2$), 2.5 ($\pm 0.5$) and 12 ($\pm 2$) $\mu$K for a range of detuning of the probe sideband from the resonance ($\Delta_r = +50, +100, +170, +230\,\Gamma$). The carrier and the blue sidebands are detuned from the resonance by $\Delta_c = \Delta_r + \Omega$ and $\Delta_b = \Delta_r + 2\Omega$, respectively. Evidently, for a fixed temperature and therefore a fixed spatial extent of the atomic ensemble, the signal initially grows linearly with atom number. Beyond a certain point in atomic density which depends on the temperature as well as the probe detuning, however, the signal deviates significantly from linearity. For high enough density, the signal ceases to be monotonic. The peak optical depth remains low (for $N = 5\times10^6$, $T = 1\mu$K and $\Delta_r = 2\pi\times 100\Gamma$, peak OD $\sim$ 0.03) accounting for a small fraction of incident light to be diffusely scattered at large angles. In addition, large probe detuning means that even at the highest peak atomic densities presented here ($\rho \sim 5 \times 10^{19}$ m$^{-3}$) the effect of light-induced dipole-dipole interaction is completely negligible \cite{Chomaz2012}.

In the diffractive model of \cite{Mueller2005, Baragiola2014}, the diffraction cone stays fixed for a fixed geometry of the atomic ensemble and the probe mode, so the mode overlap between the reference field and the scattered light fields at the detection place in the far-field should remain constant. The scattered field should thus grow linearly as a function of atom number. The nonlinear dependence of the spatial mode of the scattered light field with atom numbers, as shown by the data in Figure \ref{LensingFig4}, along with a negligible amount of diffuse, large-angle scattering, thus indicates that effects beyond the simple diffraction picture are at play. We note, in particular that at 12\,$\mu$K, the spatial extent of the atomic ensemble is well-matched to that of the probe beam and even there, the nonlinear trend is present at low detuning and large atom number.

 The nonlinear behaviour can be explained by considering atomic ensembles posing as a gradient index (GRIN) lens to the incoming light.  In Figure \ref{LensingFig4}, we also show the Fresnel number for the atomic GRIN lenses formed under the experimental conditions of the data, calculated using Equations \ref{eq:lens4} and \ref{eq:lens5}. Evidently, the departure of the dispersive signal from linearity is linked to the Fresnel number approaching 1. The linearity is maintained approximately up to an atom number that corresponds to $\mathcal{F} \sim 0.5$, shown as dashed vertical lines in Figure \ref{LensingFig4}. This, as mentioned in section \ref{GRIN_theory}, is equivalent to a ``peak phase shift'' of $\pi$ of light within the atomic medium. The $\mathcal{F} = 0.5$ line thus qualitatively separate the diffraction-dominated regime ($\mathcal{F} \lesssim 0.5$) from the lensing-dominated regime($\mathcal{F} > 0.5$).


To explain the data in Figure \ref{LensingFig4} quantitatively, we solved the paraxial wave equation for each of the propagating (scalar) electric fields \cite{SiegmanBook} corresponding to the carrier, and the two sidebands:
  \begin{equation}\label{eq:pwe}
\frac{\partial \tilde{E}}{\partial z} = \frac{i}{2k}\nabla_T^2 \tilde{E} + \frac{1}{2}i k \chi \tilde{E},
\end{equation}
where $\nabla_T^2 = \frac{\partial^2}{\partial x^2} + \frac{\partial^2}{\partial y^2}  $ is the transverse Laplacian and $\chi = \rho\alpha/\epsilon_o$ is the susceptibility of the medium, where we have assumed that $\rho\alpha \ll 1$, so the Lorentz-Lorenz correction term can be neglected. For a radially symmetric system that is not necessarily Cartesian separable, as in our case, the equation is most conveniently solved using cylindrical co-ordinates and a split-step Hankel method (Appendix). The solid lines in Figure  \ref{LensingFig4} show the dispersive signal calculated from the fields obtained from the solutions of the wave equation using the experimental parameters (number of atoms, temperature) for each data point. As is evident, the observations are in excellent agreement well the simulations. For the regime $\mathcal{F} \gtrsim 1$, the agreement become weaker which we attribute to light travelling at large angles with respect to the optical axis. This leads to uncertainties in the collected fraction of light that we did not account for, originating from, e.g., the spherical aberrations of the collecting lenses.  

\begin{figure}
        \includegraphics[width=0.985\textwidth]{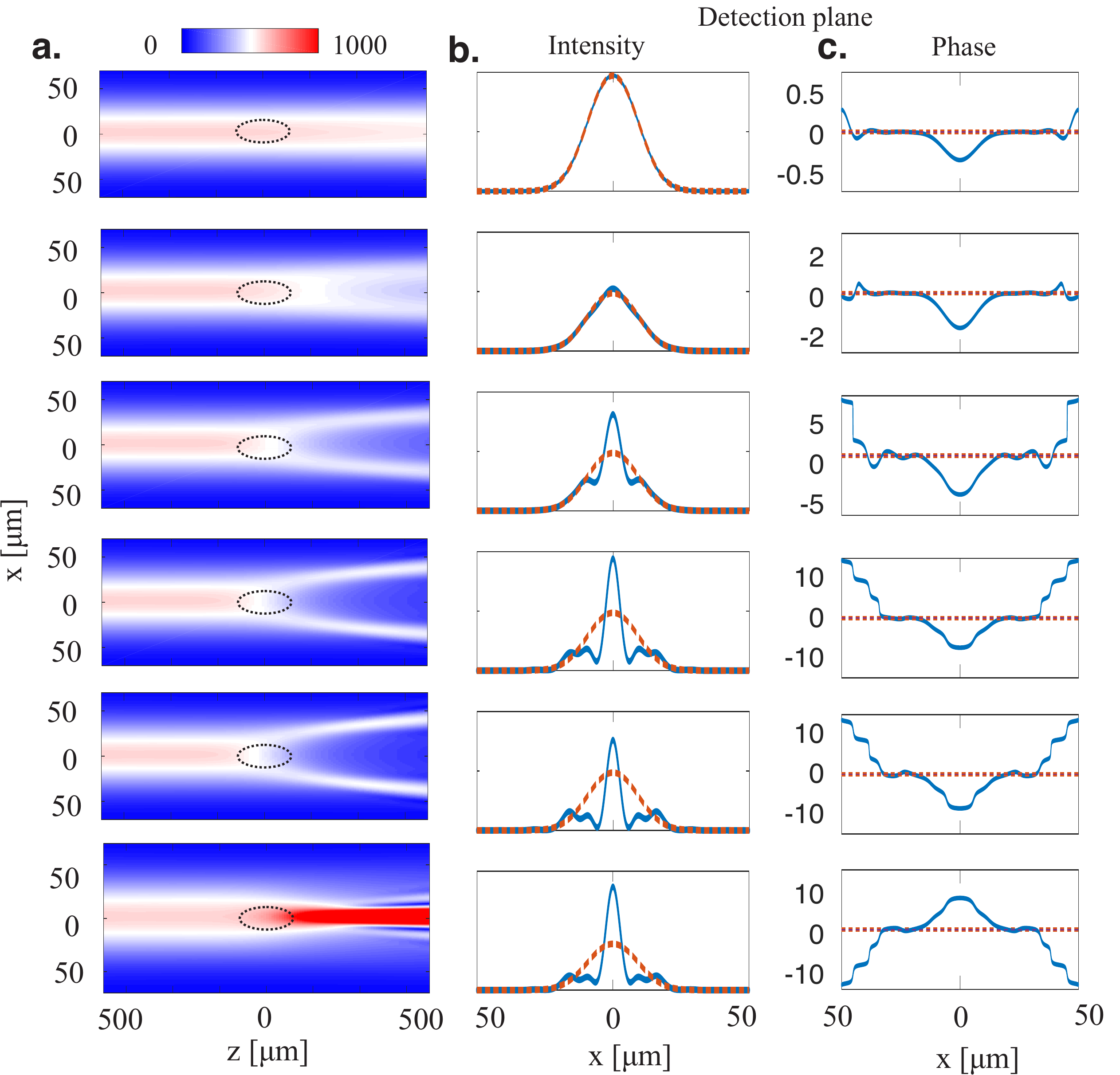}
        \vspace{-2mm}
        \caption{Transition from the diffraction-dominated regime to the lensing-dominated regime. \textbf{a.} shows the intensity distribution of the light propagating through an atomic cloud of temperature $T = 2.5 \mu$K for (top to bottom): ($N = 1\times10^5, \Delta_r = 50 \Gamma$), ($N = 5\times10^5, \Delta_r = 50 \Gamma$), ($N = 1.5\times10^6, \Delta_r = 50 \Gamma$),($N = 2.5\times10^6, \Delta_r = 50 \Gamma$). ($N = 3.5\times10^6, \Delta_r = 50 \Gamma$), and ($N = 3.5\times10^6, \Delta_r = -50 \Gamma$). The dashed ellipse shows the dimensions of the atomic cloud. \textbf{b.} and \textbf{c.}  show the correspondent intensity and phase distributions of light in the detection plane in our experiment. The dashed lines show the case with no atoms present.}
                 \label{LensingFig6}
\end{figure}

Figure \ref{LensingFig6} shows the propagation of an incident Gaussian light beam through atomic samples of varying atom numbers calculated using paraxial wave equation. For large atom numbers, light deviates significantly \textit{inside} the medium (Figure \ref{LensingFig6}a) showing that the atomic medium acts as a negative or positive lens depending on the sign of the detuning. This leads to a significant redistribution of the intensity and phase of the light field at the detection plane and a reduction of its overlap with the incident mode. The light distribution of the detected field in our experiment is at the focal plane of a lens and therefore shows the Fourier transform of the far-field distribution. The enhanced intensity peaks at the centre of the detection plane in Figure \ref{LensingFig6}b correspond to the light travelling outside the diffraction cone in the far-field (prior to the lens) due to the lensing action of the atomic ensemble.

In the far-field, the cone defined by the gaussian beam (the reference mode) is given by the angle 
  \begin{equation}\label{eq:gaussian_angle}
\theta_G = \frac{\lambda}{\pi w_0}.
\end{equation}
For small atom numbers, diffraction is the dominant effect and the scattered light stays within the diffraction cone defined by the angle
  \begin{equation}\label{eq:diffraction_angle}
\theta_D = \frac{\lambda}{2\sigma_r}.
\end{equation}
For large atom numbers, however, the atomic medium poses a strong lens bending light out of the diffraction cone. The deflection angle due to lensing is given by
  \begin{equation}\label{eq:lensing_angle}
\theta_L = \frac{\sigma_r}{f_a}.
\end{equation}
The dispersive signal, as given by Equation \ref{12} is maximised when the Gaussian cone and the diffraction cone are comparable ($\theta_G \sim \theta_D$) and so are $w_0$ and $\sigma_r$. For $ \theta_L < \theta_D$, the signal grows linearly with atom number for a fixed size. As the density grows, the focal length becomes smaller (Equation \ref{eq:lens4}) and eventually the regime $\theta_L > \theta_D$ is reached where light is scattered out of the diffraction cone and the signal in Equation \ref{12}  diminishes. Using Equations \ref{eq:lens4}, \ref{eq:diffraction_angle} and \ref{eq:lensing_angle}, we see that the condition $\theta_L > \theta_D$ is same as requiring the lens Fresnel number $\mathcal{F} \gtrsim 0.5$, which is in agreement with our observations and is corroborated by the simulation results.

\section{Summary and outlook}

We have theoretically and experimentally studied the effect of atomic lensing on the interferometric, dispersive detection of ultracold atomic ensembles in an effective zero baseline heterodyne Mach-Zender arrangement. We derived an expression for the focal length of the atomic gradient index lens and shown that the lens Fresnel number provides an useful and intuitive measure of the distinction between the diffractive and strong-lensing regimes that lead to quantitatively different dependence of the dispersive signal on the atom number. We stress that although our experimental observations are discussed in the context of a dispersive probing setup based on frequency modulation spectroscopy, both the experimental and theoretical results are relevant to any interferometric detection of coherently scattered light from an atomic ensemble. We also derive an expression for the frequency modulation spectroscopy signal for a three-dimensional interface of the sample and the probe and show that it is equivalent to zero-baseline Mach-Zender interferometers that measure the spatial-mode overlap of the coherently scattered light field with a reference field and are realized in many different ways \cite{Appel2009,Behbood2013}.

Our work augments the theoretical work on geometrical effects on dispersive detection of spin-ensembles \cite{Baragiola2014, Mueller2005} to include the effect of the atomic gradient-index lensing and experimentally demonstrate the effect. The experimental data are well-described by the wave-equation model, but more importantly, are well-supported by the qualitative criteria developed here based on atomic GRIN lens focal length and Fresnel number. This will provide an useful guide to the design choice for the atomic and probe beam geometries for dispersive detection of atomic ensembles in a wide range of applications. It is typically assumed, for instance, that for a pencil-shaped atomic sample and a probe beam of similar transverse size as the sample leads to an optimal detection of the coherently scattered field in the far-field \cite{Glauber1979, Baragiola2014}. Our analysis and experimental data shows, however, that for a given density if the length of the sample exceeds a certain point, scattered field will encroach beyond the diffraction cone. Under such conditions, the linearity of the detection signal as a function of the atom number will no longer hold. 

The complementary effect of atomic lensing is the mechanical backaction on atoms resulting from the conservation of momentum. This causes a redistribution of atomic momenta when an incident light beam is re-directed collectively by an atomic ensemble. Such optomechanical forces can lead to interesting phenomena such as electrostriction, self-structuring and effective light-induced two-body interactions between atoms \cite{Matzliah2017a,Labeyrie2013}. In the strong lensing regime, we have observed a novel geometry-dependent optomechanical self-propelling effect, which will be the subject of an upcoming publication \cite{Deb2020b}. In future studies we will explore the possible use of atomic lensing for high numerical aperture imaging and to the degree co-operative effects owing to long-range interactions can influence atomic lensing.


\section{References}

\providecommand{\newblock}{}


\section{Appendix}

\subsection*{A. Paraxial wave equation in radial co-ordinates and the Hankel Transform}

We consider the field in the cylindrical co-ordinates
  \begin{equation}\label{eq:cylind}
  E(x,y,z) \leftrightarrow u(r,\theta,z).
\end{equation}
For a field whose radial profile is of the form $u(r) \exp\{im\theta\}$, the two-dimensional Fourier transform is given by 
  \begin{equation}\label{rFFT}
\mathfrak{F}\left\{ u(r) \mathrm{r}^{im\theta}\right\} (k_r,\phi) = \mathcal{H}_m\left\{ u(r)\right\}(k_r) \mathrm{e}^{im\phi},
\end{equation}
where $\left\{ r,\theta\right\} \leftrightarrow \left\{ k_r,\phi\right\} $ are the variable pairs in the real and the transform space respectively and 
  \begin{equation}\label{Hankel1}
\mathcal{H}_m \left\{ u(r)\right\}(k_r) = \int_0^{\infty}u(r)J_m(rk_r) dr
\end{equation}
is the Hankel transform of the order $m$, where $J_m$ is the m-th order Bessel function \cite{Sigman_ch16}. 

In cylidrical co-ordinates, the transverse Laplacian operator in Equation \ref{eq:pwe} reads
   \begin{equation}\label{eq:TLp}
\nabla^2_T = \frac{\partial^2}{\partial r^2} + \frac{1}{r}\frac{\partial}{\partial r} - \frac{1}{r^2}\frac{\partial^2}{\partial\theta^2}.
\end{equation}
Substituting $E(x,y) \rightarrow u(r) e^{im\theta}$ in Equation \ref{eq:pwe} and noting that for our situation $m = 0$ (zero-fold radial symmetry), we obtain the paraxial wave equation in radial co-ordinates
 \begin{equation}\label{eq:pwe_rad}
\frac{\partial {u}}{\partial z} = \frac{i}{2k}\left ( \frac{\partial^2}{\partial r^2} + \frac{1}{r}\frac{\partial}{\partial r}  \right ) u(r) + \frac{1}{2}i k \chi(r) u(r).
\end{equation}

A split-step spectral method \cite{Press2002} can now be readily applied where field is propagated along z-axis with steps of $\delta z$ the diffraction term (first term on the right hand side of Equation \ref{eq:pwe} ) is applied for half the propagation length:
 \begin{equation}\label{eq:hn1}
 u_{j+1/2} = \mathcal{H}^{-1}\left [ \mathrm{e}^{\frac{-ik_r^2\delta_z}{4}} \mathcal{H}(u_j) \right].
\end{equation}
Here $H^{-1}$ is the inverse Hankel transform. The phase shift term is then applied
 \begin{equation}\label{eq:hn2}
 v_{j+1/2} = \mathrm{e}^{-\frac{1}{4}ik_r\chi\delta z}u_{j+1/2},
\end{equation}
which is followed by applying the diffraction term again for another half-step, yielding the field at $z = z + \delta_z$:
 \begin{equation}\label{eq:hn2}
 u_{j+1} = \mathcal{H}^{-1}\left [ \mathrm{e}^{\frac{-ik_r^2\delta_z}{4}} \mathcal{H}(v_{j+1/2}) \right].
\end{equation}

The method is computationally efficient and third-order accurate. The Hankel transform and its inverse are efficiently evaluated using a quasi-fast Hankel algorithm \cite{Siegman1976,Siegman1977}. The logarithmic radial grid used in this algorithm has the advantage that it samples densely in the region near $r = 0$, where our atomic density is the highest and sparsely far-away from the paraxial region. This, combined with the one-dimensional nature of the problem, allows us to evaluate the field over a large transverse region and propagate it directly to the far-field through external lens elements and apertures.

\end{document}